\documentclass[prd,showpacs,showkeys,preprint]{revtex4}

\usepackage{amsmath}
\usepackage{amssymb}
\usepackage{yfonts}[1988/10/03]
\usepackage{graphicx}
\newcommand {\beq}{\begin{equation}}
\newcommand {\eeq}{\end{equation}}
\newcommand {\beqa}{\begin{eqnarray}}
\newcommand {\eeqa}{\end{eqnarray}}

\begin{document}
\title{The investigation of detectability of the relic gravitational waves based on the WMAP-9 and Planck }% Force line breaks with \\
\author{ Basem Ghayour$^{1}$\footnote{ba.ghayour@gmail.com}}
\affiliation{$^{1}$ School of Physics, University of Hyderabad,
Hyderabad-500 046. India.
}
%\affiliation{$^{(b)}$Department of Physics, Isfahan University of Technology, Isfahan 84156-83111, Iran}
%\maketitle

\date{\today} % It is always \today, today,

\begin{abstract}
The generated relic gravitational waves were underwent several stages of
evolution of the universe such as inflation and reheating. These stages  were affected on the  shape of  spectrum of
the waves. As well known, at the end of inflation, the scalar field $\phi$ oscillates quickly around some
point where potential $V(\phi)=\lambda \phi^{n}$ has a minimum. The end of inflation stage played
a crucial role on the further evolution stages of the universe because particles were
created and collisions of the created particles were responsible for reheating the universe. There is a general range for the frequency of the spectrum $ \sim (0.3\times10^{-18}-0.6\times10^{10}$)Hz. It is shown that the reheating temperature  can
be affect on the frequency of the spectrum as well. There is constraint on the temperature
from cosmological observations based on WMAP-9 and Planck. Therefore it is interesting to estimate allowed value of frequencies of the spectrum based on general
range of reheating temperature like few MeV $\lesssim T_{rh}\lesssim10^{16}$ GeV, WMAP-9 and Planck data then compare  the spectrum with sensitivity of
future detectors such as LISA, BBO and ultimate-DECIGIO. The obtained results of this comparison give us some more chance for detection of the relic gravitational waves.
\end{abstract}
\maketitle
\section{\label{sec:level1}Introduction }
The relic gravitational waves generated  in the early universe are very importance  because they provide useful information about  physics of  the early  universe. The waves  were underwent  several stages of  evolution of the  universe such as: inflation, reheating, radiation, matter and acceleration. The  evolution of  the   universe   at various epoch is    affected the  shape of  the spectrum of the waves.  Therefore  it is unavoidable  the consideration of different stages of  evolution of the universe on  the study of  spectrum of the waves   that are  to be observed  today.   It is believed that the universe underwent a quasi exponential  expansion in its early stages of evolution known as inflation stage. This stage is  
very important   on the evolution  of the universe  because   it provided  the mechanism to formation of large scale structures in the universe. At the end of inflation, the scalar field  $\phi$ oscillates quickly around some
point where potential $V(\phi)=\lambda \phi^{n}$ has a minimum.   The  end of  this stage  played a  crucial  role  on the further evolution  stages of the universe because  particles were created and  collisions of the created particles were responsible  for   reheating  the universe.  The reheating was essential for  the nucleosynthesis process since the inflation  brought  temperature of the universe below  for the  requirement of thermo nuclear reactions. Towards the end of inflation, during the reheating, the equation of state of
energy for the universe is   quite complicated and also model-dependent \cite{qa}. Hence  a new stage that called 
$z$-stage is introduced to allow a general behaviour of reheating epoch \cite{3}.  

There is a general range for the frequency of the spectrum $\sim$($ 0.3\times10^{-18}-0.6\times10^{10}$) Hz. 
  It is shown that  the  reheating temperature ($T_{rh}$) can be affect on the  frequency of the spectrum  \cite{acq}. There is  constraint on the $T_{rh}$    from cosmological observations based on WMAP-9 and Planck.   However, the reheating temperature  must be larger than a few MeV \cite{qw}, for the creation light elements,
but less than the  energy scale at the end of inflation,  that is $T_{rh}\lesssim10^{16}$ GeV.   Therefore it is interesting to estimate allowed value of frequencies of the spectrum based on general range of $T_{rh}$, WMAP-9 and Planck data then compare the spectrum with sensitivity of future detectors such as LISA \cite{nb}, Big Bang Observer (BBO) \cite{aa} and   ultimate DECI-hertz Interferometer Gravitational-wave Observatory (ultimate -DECIGO) \cite{aas}. Hence the main purpose of this work is investigation of   this comparison. The obtained results of this comparison give us some more chance for detection of the waves. We use the units $c=\hbar=k_{B}=1$.

\section{Gravitational waves  spectrum in expanding universe}\label{two}
The perturbed metric for a homogeneous  isotropic  flat Friedmann-Robertson-Walker  universe  can be written   as
\begin{equation}
d s^{2}= S^{2}(\eta)(d\eta^{2}-(\delta_{ij}+h_{ij})dx^{i}dx^{j}),
\end{equation}
where $S(\eta)$ is the cosmological scale factor,  $\eta$ is the conformal time and $\delta_{ij}$ is the Kronecker delta  symbol.  The $h_{ij} $ are  metric perturbations  field contain only the pure gravitational waves and are  transverse-traceless i.e; $\nabla_i h^{ij} =0, \delta^{ij} h_{ij}=0$.

The present study  consider  the shape of  the spectrum of  relic gravitational waves  that generated by the expanding space time  background. Thus  the perturbed matter source is  not taken into account. The 
 gravitational waves are described with  the  
 linearized field equation given by 
\begin{equation}\label{weq}
\nabla_{\mu} \left( \sqrt{-g} \, \nabla^{\mu} h_{ij}(\bf{x}, \eta)\right)=0.
 \end{equation} 
The tensor perturbations have two independent physical degrees of freedom like $h^{+}$ and $h^{\times}$ and called polarization modes. To compute the spectrum of gravitational waves $h(\bf{x},\eta)$, we express $h^{+}$ and $h^{\times}$ in terms
of the creation ($a^{\dagger}$) and annihilation ($a$) operators,
\begin{equation}\label{1q1}
 h_{ij}(\textbf{x},\eta)=\dfrac{\sqrt{16\pi} l_{pl}}{S(\eta)} \sum_\textbf{p} \int\dfrac{d^{3}k}{(2\pi)^{3/2}} {\bf \epsilon}_{ij} ^{\textbf{p}}(\textbf {k}) 
 \times  \frac{1}{\sqrt{2 k}} [a_{\textbf{k}}^{\textbf{p}}h_\textbf{k}^{\textbf{p}}(\eta) e^{i\textbf{k}.\textbf{x}} +a^{\dagger}_{\textbf{k}} {^{\textbf{p}}} h^{*}_\textbf{k}{^{\textbf{p}}} (\eta)e^{-i\textbf{k}.\textbf{x}}],
\end{equation}
where  $\bf{k}$ is the comoving wave
number, $k=|\bf {k}|$, $l_{pl}= \sqrt{G}$ is the
Planck's length and $\bf{ p}= +, \times$ are polarization modes. The polarization tensor 
$\epsilon_{ij} ^{{\bf p}}({\bf k})$ is symmetric and transverse-traceless  $ k^{i} \epsilon_{ij} ^{{\bf p}}({\bf k})=0, \delta^{ij} \epsilon_{ij} ^{{\bf p}}({\bf k})=0$ and 
satisfy  the conditions $\epsilon^{ij {\bf p}}({\bf k})   \epsilon_{ij}^{{\bf p}^{\prime}}({\bf k})= 2  \delta_{ {\bf p}{{\bf p}}^{\prime}} $ and $ \epsilon^{{\bf p}}_{ij} ({\bf -k}) = \epsilon^{{\bf p}}_{ij} ({\bf k}) $. The $a$ and $a^{\dagger}$   satisfy
$[a_{{\bf k}}^{{\bf p}},a^{\dagger}_{{\bf k} ^{\prime}} {{^{{\bf p}}}^{\prime}}]= \delta_{{{\bf p}} {\bf {p}}^{\prime} }\delta^{3}({\bf k}-{{\bf k}}^{\prime})$ and the initial vacuum state is defined  as $a_{\bf{k}}^{\bf{p}}|0\rangle = 0$
%\begin{equation}
%a_{\bf{k}}^{\bf{p}}|0\rangle = 0,
%\end{equation}
for each $\bf {k}$ and $\bf {p}$.

For a fixed  wave number $\bf{k} $ and a fixed polarization state $\bf{p}$ the linearized wave eq.(\ref{weq}) gives
 \begin{equation}\label{zz1}
h^{\prime \prime}_{k}+2\frac{S^{\prime}}{S}h^{\prime}_{k}+k^{2}{h}_{k}=0,
\end{equation}
where   prime means derivative with respect to the conformal time. Because the  polarization states are  same,   we consider   $h_{k}(\eta)$ without the polarization  index.  

Next, we  rescale the filed $h_{k}(\eta)$ by taking
$h_{k}(\eta)=f_{k}(\eta)/S(\eta)$, where the mode functions $f_{k}(\eta)$ obey the minimally coupled Klein-Gordon equation
\begin{equation}\label{zz}
f^{\prime \prime}_{k}+\Big(k^{2}-\frac{S^{\prime \prime}}{S} \Big)f_{k}=0.
\end{equation}

The general solution of the above equation is a linear combination of the Hankel function with
a generic power law for the scale factor $S=\eta^{q}$ given by

\begin{equation}\label{uy}
f_{k}(\eta)=A_{k}\sqrt{k\eta}H^{(1)}_{q-\frac{1}{2}}(k\eta)+B_{k}\sqrt{k\eta}H^{(2)}_{q-\frac{1}{2}}(k\eta).
\end{equation}

For a given model of the  universe, consisting of a sequence of successive scale
factors with different $q$,  we can obtain an exact solution $f_{k}(\eta)$ by matching its value and
derivative at the joining points.

The  approximate computation of the spectrum   is  calculated in two cases depending up on the waves that are  outside  or within of the  barrier. For the  gravitational  waves  outside barrier ($k^{2}\gg S^{\prime \prime}/S$)  the corresponding   amplitude  decrease as $h_k \propto 1/S(\eta) $ and for the  waves inside the barrier ($k^{2} \ll S^{\prime \prime}/S$),  $h_k = C_k $ simply a constant \cite{po}.

The history of   expansion of the universe can be obtained as  follows: 
The inflation stage 
\begin{equation}\label{p}
S(\eta)=l_{0}|\eta |^{1+\beta},\;\;\;\;\;\;-\infty <\eta\leq \eta_{1},
\end{equation}
where $1+\beta <0$, $\eta<0$ and $l_{0}$ is a constant.

To
make our analysis more general, we consider that the inflation stage was followed by some interval
of the z-stage (z from Zeldovich). In fact this stage is quite general that considered by Zeldovich \cite{aqz}.  It can be governed by a softer than radiation matter, as well as by a stiffer than radiation matter  \cite{de}.  Towards the end of inflation, during the reheating, the equation of state of
energy in the universe can be quite complicated  \cite{qa}. Hence this
z-stage is introduced to allow a general reheating stage. Therefore we define the reheating stage in general form
\begin{equation}\label{pq}
S(\eta)=S_{z}(\eta - \eta_{p})^{1+\beta_{s}},\;\;\;\;\;\;\eta_{1}<\eta\leq \eta_{s},
\end{equation}
where $1+\beta_{s}>0$ \cite{po}. 

The radiation-dominated stage
\begin{equation}
S(\eta)=S_{e}(\eta-\eta_{e}),\;\;\;\;\;\;\eta_{s}\leq \eta \leq \eta_{2},
\end{equation}
and the matter-dominated stage
\begin{equation}
S(\eta)=S_{m}(\eta-\eta_{m})^{2},\;\;\;\;\;\;\eta_{2}\leq \eta \leq \eta_{E},
\end{equation}
where $\eta_{E}$ is the time when the dark energy density $\rho_{\Lambda}$ is equal to the matter energy density $\rho_{m}$. 

The value of  redshift $z_{E}$ at  $\eta_{E}$ is  $(1+z_{E})=S(\eta_{0})/S(\eta_{E})\sim 1.3 $ from Planck collaboration \cite{ew1}, where $\eta_{0}$ is the present time.  

The accelerating stage (up to the present)
\begin{equation}\label{1w}
S(\eta)=\ell_{0}|\eta- \eta_{a} |^{-\gamma},\;\;\;\;\;\;\eta_{E}\leq \eta \leq\eta_{0},
\end{equation}
where $\gamma$ is $\Omega_{\Lambda}$ dependent parameter, and $\Omega_{\Lambda}$ is the energy density contrast. We take  $\gamma\simeq 1.97$ \cite{mn} for $\Omega_{\Lambda}=0.73$ \cite{mb}.

 Except for  $\beta_{s}$, there are ten
constants in the  expressions of $S(\eta)$. By the continuity conditions of $S(\eta)$ and $S^{\prime}(\eta)$ at
 four given joining points $\eta_{1}, \eta_{s}, \eta_{2},$ and $\eta_{E}$, one can fix only eight constants. The remaining
two constants can be fixed by the normalization of $S$ and  the observed Hubble
constant. We put $|\eta_{0}-\eta_{a}|=1$ for the normalization of $S$, which fixes the  $\eta_{a}$, and the constant $\ell_{0}$ is fixed by the following calculation,
\begin{equation}
\frac{\gamma}{H}\equiv \Big(\frac{S^{2}}{S^{\prime}}\Big)_{\eta_{0}}=\ell_{0},
\end{equation}
where $\ell_{0}$ is  the Hubble radius at present and $H=100 \;h$ km s$^{-1}$ Mpc$^{-1}$ with $h\simeq 0.704$ \cite{mb}. 

The physical wavelength is related to the comoving wave
number  as
$\lambda \equiv 2\pi S(\eta)/k$. Assuming that
the wave mode crosses the horizon of the universe when $\lambda/2\pi=1/H$ \cite{ws},
then wave number $k_{0}$ corresponding to the present Hubble radius is
$k_{0}= S(\eta_{0})/ \ell_{0}= \gamma$. Also
there is another wave number
$k_{E}=\frac{ S(\eta_{E})}{1/H}=\frac{k_{0}}{1+z_{E}},$
whose corresponding wavelength at the time $\eta_{E}$ is the Hubble radius $1/H$.

By matching $S$ and $S^{\prime}/S$ at the joint points, one gets
\begin{equation}\label{kk}
l_{0}=\ell_{0}b\zeta_{E}^{-(2+\beta)}\zeta_{2}^{\frac{\beta-1}{2}}\zeta_{s}^{\beta}\zeta_{1}^{\frac{\beta-\beta_{s}}{1-\beta_{s}}},
\end{equation}
where $b\equiv|1+\beta|^{-(2+\beta)}$, $\zeta_{E}\equiv\frac{S(\eta_{0})}{S(\eta_{E})}$, $\zeta_{2}\equiv\frac{S(\eta_{E})}{S(\eta_{2})}$, $\zeta_{s}\equiv\frac{S(\eta_{2})}{S(\eta_{s})}$, and $\zeta_{1}\equiv\frac{S(\eta_{s})}{S(\eta_{1})}$. 

The power spectrum  of   gravitational waves is defined  as
\begin{equation}\label{pow}
\int_0 ^\infty h^2 (k,\eta) \frac{dk} {k} = \langle 0 | h^{ij}({\bf x},\eta) h_{ij}({\bf x},\eta) |0 \rangle .
\end{equation}
 Substituting eq.(\ref{1q1}) in eq.(\ref{pow}) and taking the contribution from each polarization is same, we get
 \begin{equation}\label{pp}
 h(k,\eta)= \frac{4 l_{pl}}{\sqrt{\pi}} k | h(\eta)|.
 \end{equation} 
Thus once the mode function $h(\eta)$ is known, the spectrum $h(k,\eta)$ follows.

The spectrum at the present time $ h(k,\eta_0)$ can be obtained, provided the initial  spectrum is specified. The initial condition is taken to be during the inflation. The wave with wave number $k$ crossed over the horizon at a time $\eta_{i}$, when the wavelength $\lambda_{i}/2\pi =1/H(\eta_{i})=S(\eta_{i})/k$ \cite{ws}. Now we choose the initial condition of the mode
function $h_{k}$ as $|h_{k}(\eta_{i})|=1/S(\eta_{i})$. The initial amplitude of the power spectrum is
\begin{equation}\label{vb}
h(k,\eta_{i})=8\sqrt{\pi}\frac{l_{pl}}{\lambda_{i}}.
\end{equation}
With $\lambda_{i}/ 2\pi=1/H(\eta_{i})$ it becomes
\begin{equation}\label{vn}
\frac{S^{\prime}(\eta_{i})}{S(\eta_{i})}=k.
\end{equation}
 Therefore  initial amplitude of the spectrum is given by
\begin{equation}\label{bet}
h(k,\eta_i)= A{\left(\frac {k}{k_0}\right)}^{2+\beta},
\end{equation}
where the constant $A$ in eq.(\ref{bet}) can be determined by quantum normalization \cite{po}: 
 \begin{equation}\label{bet1}
 A=8\sqrt{ \pi} \frac{l_{pl}}{l_0}. 
  \end{equation}

 Thus the amplitude  of the spectrum for  different ranges are given as follows \cite{po}, \cite{15}, \cite{vb}, \cite{lk}.

\begin{equation}\label{y}
h(k,\eta_{0})=A\Big(\frac{k}{k_{0}}\Big)^{2+\beta},\;\;\;k\leq k_{E},
\end{equation}

\begin{equation}\label{ke}
h(k ,\eta_{0})=A\Big(\frac{k}{k_{0}}\Big)^{\beta-\gamma} (1+z_{E})^{\frac{-2-\gamma}{\gamma}},\;\;\;k_{E}\leq k\leq k_{0},
\end{equation}

\begin{equation}\label{l}
h(k,\eta_{0})=A\Big(\frac{k}{k_{0}}\Big)^{\beta} (1+z_{E})^{\frac{-2-\gamma}{\gamma}},\;\;\;k_{0}\leq k\leq k_{2},
\end{equation}

\begin{equation}\label{o}
h(k,\eta_{0})=A\Big(\frac{k}{k _{0}}\Big)^{1+\beta}\Big(\frac{k_{0}}{k_{2}}\Big)(1+z_{E})^{\frac{-2-\gamma}{\gamma}},\;\;\;k _{2}\leq  k \leq k _{s},
\end{equation}

\begin{equation}\label{oo}
h(k,\eta_{0})=A\Big(\frac{k}{k _{0}}\Big)^{1+\beta-\beta_{s}}\Big(\frac{k _{s}}{k _{0}}\Big)^{\beta_{s}}\Big(\frac{k _{0}}{k _{2}}\Big)(1+z_{E})^{\frac{-2-\gamma}{\gamma}},\;\;\;k _{s}\leq k \leq k _{1}.
\end{equation}

The  factor $A$  in all the spectra is determined   with the CMB data of  WMAP-9 \cite{115}.
  The observed CMB anisotropies at lower multipoles is $\Delta T / T \simeq0.44\times10^{-5}$ at $l\sim2$ which corresponds to the largest
scale anisotropies that have observed so far. Thus one can gets
\begin{equation}\label{k}
h(k_{0},\eta_{0})=A(1+z_{E})^{\frac{-2-\gamma}{\gamma}}\simeq 0.44 \times 10^{-5}r.
\end{equation}
where $r$  is tensor to scalar ratio \cite{q2} (see the appendix.(A)  for more details). The parameter $r$ taken $\sim0.1$ from Planck collaboration \cite{ew1}.
However, there is a point in the interpretation of $\delta T/T$ at low multipoles. At present, the Hubble radius and Hubble
diameter are $\ell_{0}$,  $2\ell_{0}$ respectively. 
The corresponding physical wave length of $k_{E}$ at present is $ S(\eta_{0})/k_{E}=\ell_{0}(1+z_{E})\simeq 1.32 \ell_{0}$,  which
is within $2\ell_{0}$ and is theoretically observable. So, instead of eq.(\ref{k}), if $\delta T/T \simeq 0.44 \times 10^{-5}$ at $l=2$ were taken as the amplitude of the spectrum at $k_{E}$, then we have $h(k_{E}, \eta_{0})=0.44 \times10^{-5}\times r^{1/2}$  yielding a smaller $A$ than
that in eq.(\ref{k}) \cite{q3,ad}. Also there is another normalization method that leads to decaying factor $S(\eta_{i})/S(\eta_{0})$ \cite{mna}. 

 By taking   $\nu$ as frequency, we can obtain $\nu_{E}=0.30\times10^{-18}$ Hz, $\nu_{0}=0.36\times10^{-18}$ Hz, $\nu_{2}=1.48\times10^{-17}$ Hz, $\nu_{s}=0.15\times10^{8}$ Hz.

The spectral energy density
parameter $\Omega_{g}(\nu)$ of gravitational waves is defined through the relation $\rho_{g}/\rho_{c}=\int\Omega_{g}(\nu)\frac{d\nu}{\nu}$, where $\rho_{g}$ is the energy density of the gravitational waves and $\rho_{c}$ is the critical energy density.
Therefore we have \cite{po}
\begin{equation}\label{ka}
\Omega_{g}(\nu)=\frac{\pi^{2}}{3}h^2(k,\eta_{0})\Big(\frac{\nu}{\nu_{0}}\Big)^{2}.
\end{equation}

 We assume that the   space time is  spatially flat 
$K=0$ with $\Omega=1$, then the fraction density of relic gravitational waves must be less than unity, $\rho_{g}/ \rho_{c}<1$. In order to $\rho_{g}/ \rho_{c}$ dose not exceed the level of  $10^{-5}$, the $\Omega_{g}(\nu_1)\simeq 10^{-6}$ in eq.(\ref{ka})  therefore we get $\nu_1\simeq 3\times10^{10} $ Hz \cite{po}. When the acceleration epoch is considered,
the constraint becomes $\nu_1\simeq 4\times10^{10}$ Hz \cite{acq}. 

So far we did not take the effect of reheating temperature on the spectrum of gravitational waves. Then we will consider this effect in the next section.

\section{The effect of reheating temperature on the spectrum}\label{yy}

At the end of inflation, 
the scalar field $\phi$ oscillates quickly around some point where potential $V(\phi)$ has a minimum. It is found that the scalar field
oscillations behave like a fluid with $p=\omega\rho$, where the average equation of state $\omega$ depends
on the form of the potential $V(\phi)$ \cite{aq}. For $V(\phi)=\lambda\phi^{n}$, one has
\begin{equation}\label{qw}
\omega =\frac{n-2}{n+2}.
\end{equation}
There is theoretical consideration during inflation and reheating stages for the equation of state -1/3 $<\omega<$ 1 \cite{acq}. Due to eq.(\ref{qw}) the condition  leads to $n>1$. The upper bound based on CMB observation for $n$ gives  $n < 2.1$ \cite{fv}. Therefore, we can write the range of $n$ as follows
\begin{equation}\label{cc}
1<n<2.1.
\end{equation}
There are two relations that connect the $\beta$ and $\beta_{s}$ with $n$ \cite{acq}:
\begin{equation}\label{ca}
\beta_{s}=\frac{4-n}{2(n-1)},
\end{equation}
and
\begin{equation}\label{bn}
\beta=-2-\frac{n}{2(n+2)}(1-n_{s}),
\end{equation}
 where $n_{s}$ is scalar spectral index. And also we can write the $T_{rh}$ as follows \cite{bn}
\begin{equation}\label{uy}
T_{rh}=3.36\times10^{-68}\sqrt{\frac{1-n_{s}}{A_{s}}}\;exp(\frac{6}{1-n_{s}}),
\end{equation}
where $A_{s}$ is  amplitude of the scalar perturbations.  For taking in account the effect of the $T_{rh}$ on the spectrum, we can consider the following relations \cite{acq,po}: 
\begin{equation}\label{wq}
\zeta_{s}=\Big( \frac{\nu_{s}}{\nu_{2}}\Big) =\frac{S(\eta_{2})}{S _{rec}}\frac{S_{rec}}{S(\eta_{s})}=\frac{T_{rh}}{T_{CMB}(1+z_{eq})}\Big(\frac{g_{1}}{g_{2}}
\Big )^{1/3},
\end{equation}

\begin{equation}\label{ew}
\zeta_{1}=\Big( \frac{\nu_{1}}{\nu_{s}}\Big)^{(1+\beta_{s})} =
\frac{S(\eta_{s})}{S (\eta_{1})}=\frac{m_{pl}}{k^{p}_{0}}\Big[\pi A_{s}(1-n_{s})\frac{n}{2(n+2)}  \Big]^{1/2}\frac{T_{CMB}}{T_{rh}}\Big (  \frac{g_{2}}{g_{1}} \Big)^{1/3} exp \Big [- \frac{n+2}{2(1-n_{s})} \Big],
\end{equation}

 \begin{table}[t]
{ The obtained $T_{rh}$ is  based on WMAP-9. The $n_{s}$, $A_{s}$ and $m$ are taken  from \cite{ew} and $W$  stands for WMAP.}
{\begin{tabular}{@{}ccccc@{}} \toprule
$Object$&$
n_{s}$&$\;\;A_{s}\times10^{9}$&$m\times10^{16} (GeV)$ \;&$T_{rh} (GeV)$\\ \hline
$W$&$ 0.972\pm0.013$&\;\;$2.41\pm0.10$ &  $1.05\;m_{pl}$ &   $1.324\times10^{29}$ \\\\
$W+eCMB$&$0.9642\pm0.0098$&$\;\;2.43\pm0.084$ & $1.35\;m_{pl}$ &   $7.8\times10^{8}$\\\\
$W+eCMB+BAO$&$ 0.9579^{+0.0081}_{-0.0082}$&$\;\;2.484^{+0.073}_{-0.072}$ &  $1.61\;m_{pl}$ &   $0.0108$ \\\\
$W+eCMB+H_{0}$&$ 0.9690^{+0.0091}_{-0.0090}$&$\;\;2.396^{+0.079}_{-0.078}$ &  $1.16\;m_{pl}$ &   $1.378\times10^{20}$ \\\\
$W+eCMB+BAO+H_{0}$\;&$ 0.9608\pm 0.008$\;&$\;\;2.464\pm 0.072$ \;&  $1.49\;m_{pl}$ &   $399.171$\\
\\ \botrule
\end{tabular} \label{sa}}
\end{table}

 \begin{table}[t] 
{ The obtained  $T_{rh}$ is based on Planck. The $n_{s}$, $A_{s}$ and $m$ are taken  from \cite{ew1}. The $P$, $le$, $W$, and $hl$  stand for Planck, lensing,  WMAP and HighL respectively.}
{\begin{tabular}{@{}ccccc@{}} \toprule
$Object$&$n_{s}$&$\;\;ln(10^{10}A_{s})$&$\;m\times10^{16} (GeV)$ &$\;T_{rh} (GeV)$\\ \hline
$P$&$ 0.9616\pm0.0094$&$\;\;\;3.103\pm0.072$ &  $\;\;\;\;1.39\;m_{pl}$ &   $1.007\times10^{4}$ \\\\
$P+le$&$0.9635\pm0.0094$&$\;\;\;3.085\pm0.057$ & $\;\;\;\;1.30\;m_{pl}$ &   $3.4\times10^{7}$\\\\
$P+W$&$ 0.9603\pm0.0073$&$\;\;\;3.089^{+0.024}_{-0.027}$ &  $\;\;\;\;1.42\;m_{pl}$ &   $61.867$\\\\
$P+W+hl$&$ 0.9585\pm0.007$&$\;\;\;3.090\pm0.025$ &  $\;\;\;\;1.49\;m_{pl}$ &   $0.090$\\\\
$P+le+W+hl$&$ 0.9641\pm0.0063$&$\;\;\;3.087\pm0.024$ &  $\;\;\;\;1.38\;m_{pl}$ &   $4.531\times10^{3}$\\\\
$P+W+hl+BAO$&$0.9608\pm0.0054$&$\;\;\;3.091\pm0.025$ &  $\;\;\;\;1.41\;m_{pl}$ &   $422.196$\\
 \botrule
\end{tabular}\label{sas1}}
\end{table}
   
where $S_{rec}$ and $T_{rec}$ stand for the scale factor and the temperature at the recombination,
respectively and $k^{p}_{0}=0.002$ Mpc$^{-1}$ is pivot wavenumber. The $g_{1}=200$ and $g_{2}=3.91$ count the effective number of relativistic species contributing to the
entropy during the reheating and that during recombination respectively \cite{acq}. Also we used $T_{rec}=T_{CMB}(1+z_{rec})$ with $T_{CMB}=2.725$ K $=2.348\times10^{-13}$ GeV \cite{mb}. 

Using eq.(\ref{uy}), the obtained $T_{rh}$ for given $n_{s}$ and $A_{s}$ based on different types of the object for WMAP-9 (WMAP-9 +eCMB, WMAP-9 +eCMB+BAO, ...) and Planck (Planck+lensing, Planck+WMAP-9, ...) are shown in tables.(\ref{sa},\ref{sas1}) respectively.  Therefore the  acceptable range for $T_{rh}$ corresponds to the range   few MeV $\lesssim T_{rh}\lesssim10^{16}$ are obtained as follows

\begin{equation}\label{ui}
0.0108 \leq T_{rh} \leq 7.8\times10^{8}\;\;\; GeV,\;\;\;\;\;\;WMAP-9,
\end{equation}

\begin{equation}\label{oi}
0.090\leq T_{rh} \leq 3.4\times10^{7}\;\;\;\;\; GeV,\;\;\;\;\;\;Planck.
\end{equation}

Then we   find the frequencies $\nu_{s}$ and $\nu_{1}$ as function of the $T_{rh}$ with help of eqs.(\ref{wq}$-$\ref{oi}). The   obtained frequencies are smaller than  their initial amount ($\nu_{s}=0.15\times10^8$ Hz, $\nu_{1}=4\times10^{10}$ Hz) due to $T_{rh}$ as follows

\begin{equation*}
0.78\times10^{-9}\leq \nu_{s}\leq 0.55\times10^{2}\;\;\;Hz,\;\;\; with \;\;\;\;n\sim 1,
\end{equation*}
\begin{equation}\label{w}
0.47\times10^{-7}\leq \nu_{1}\leq 0.44\times10^{3}\;\;\;Hz ,\;\;\; with \;\;\;\;n\sim 1,
\end{equation}

\begin{equation*}
0.78\times10^{-9}\leq \nu_{s}\leq 0.55\times10^{2}\;\;\;Hz,\;\;\; with\;\;\; n\sim 2.1,
\end{equation*}
\begin{equation}\label{q}
0.28\times10^{3}\leq \nu_{1}\leq 1.16\times10^{6}\;\;\;\;\;Hz,\;\;\; with\;\;\; n\sim 2.1,
\end{equation}
for WMAP-9 and

\begin{equation*}
0.65\times10^{-8}\leq \nu_{s}\leq 2.45\times10^{0}\;\;\;\;Hz,\;\;\;with\;\;\;n\sim 1,
\end{equation*}
\begin{equation}\label{a}
0.33\times10^{-6}\leq \nu_{1}\leq 0.23\times10^{2}\;\;\;\;\;Hz,\;\;\;\;with\;\;\;n\sim 1,
\end{equation}

\begin{equation*}
0.65\times10^{-8}\leq \nu_{s}\leq 2.45\times10^{0}\;\;\;Hz,\;\;\;\;with\;\;\;n\sim 2.1,
\end{equation*}
\begin{equation}\label{s}
0.57\times10^{3}\leq \nu_{1}\leq 0.39\times10^{6}\;\;\;\;\;\;Hz,\;\;\;\;\;with\;\;\;n\sim 2.1,
\end{equation}
for Planck. It is noted that the $T_{rh}$ does not change the frequencies less than $\nu_{s}$ and $\nu_{1}$. 

\begin{figure}
 {\includegraphics[scale=0.5]{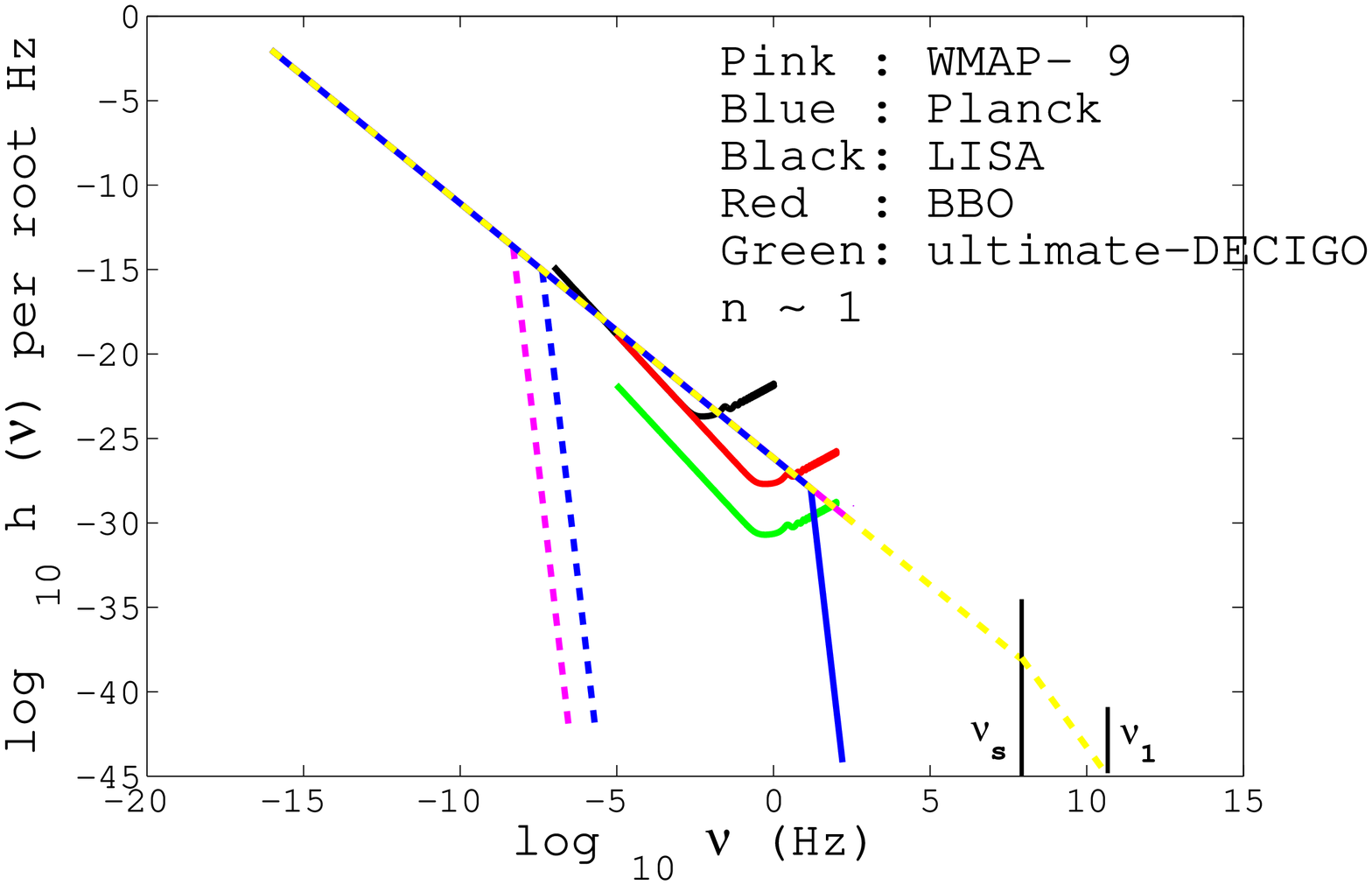}}
\caption{\label{ff1}The  comparison of the spectrum of gravitational waves  with the sensitivity of detectors such as LISA, BBO and ultimate-DECIGO for $n\sim 1$.  }
\end{figure}

\begin{figure}[]
 {\includegraphics[scale=0.5]{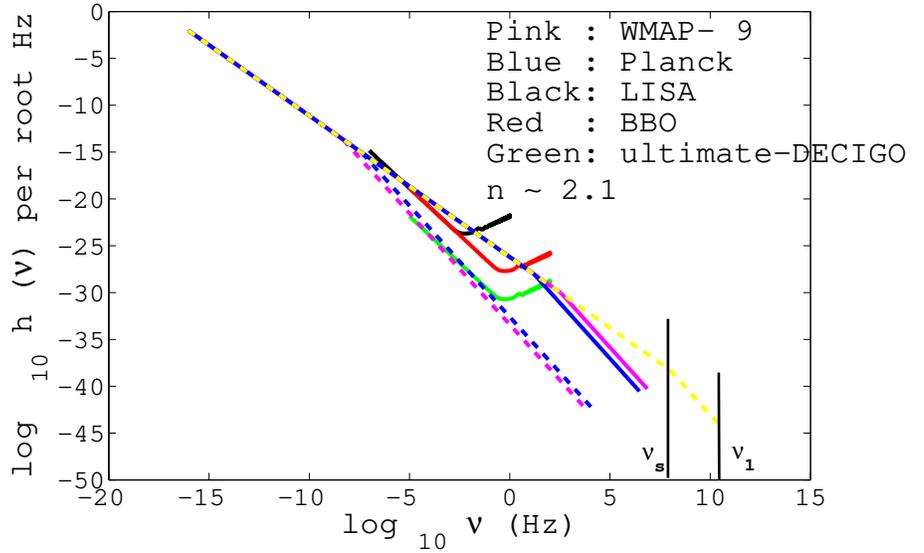}}
\caption{\label{ff2}The  comparison of the spectrum of gravitational waves  with the sensitivity of detectors such as LISA, BBO and ultimate-DECIGO for $n \sim 2.1$. }
\end{figure}

The obtained frequencies based on WMAP and Planck can affect on the shape of the spectrum of the waves  in the range $\nu_{s}$ up to  $\nu_{1}$ and gives us interesting results about  detection of the waves. We plot the spectrum in Figs.[\ref{ff1}, \ref{ff2}]  compared to the  sensitivity of detectors  such as LISA, BBO and ultimate-DECIGO (black, red and green colors respectively). The yellow, pink and blue colors are based on the initial amount of ($\nu_{s}, \nu_{1}$),  WMAP-9 and Planck data respectively.  Moreover the dashed  (solid) of pink and blue colors are based on the  minimum (maximum) amount of ($\nu_{s}, \nu_{1}$) due to $T_{rh}$ from  eqs.(\ref{ui}, \ref{oi}) in both figures. It is noted that there are some overlap in dashed yellow and  solid blue  for both figures.

It is shown that all detectors can detect the waves corresponds to  the amount of $T_{rh}$ in both figures. But, there is no chance for detection of the waves  due to minimum amount of $T_{rh}$ (dashed pink and blue colors) with all detectors  while  it exists for the maximum  amount  of $T_{rh}$  for the ultimate-DECIGO (solid pink and blue colors) in Figs.[\ref{ff1}, \ref{ff2}].  The same thing has happened about the detection of the waves  due to minimum amount of $T_{rh}$ for LISA and BBO in Fig.[\ref{ff2}].   But there is some more chance about the detection of the waves in full range  of the $T_{rh}$ with ultimate DECIGO in Fig.[\ref{ff2}]. It is noted that there are some other methods for the detection of the waves at very high frequency range $\nu_{s}$ up to $\nu_{1}$ based on wave guide \cite{az} and Gaussian beam
\cite{azz}.

Therefore based on this work, it is shown that the reheating era and reheating temperature play main role in the shape and evolution of the waves in the range $\nu_{s}$ up to  $\nu_{1}$. Hence there is no reason for looking for the detection of gravitational waves in the  high range (more than  $10^{3}$Hz and $10^{6}$Hz correspond to the eqs.(\ref{w}-\ref{s})). It is observed that there are some intersections between the spectrum and sensitivity of the detectors especially ultimate DECIGO in the mentioned range. Hence we hope to detect the waves with future mentioned mission of the detectors.

\section{Discussion and conclusion}
  
The relic gravitational waves that originated in the early universe are very important in cosmology because they  carry information about the physical conditions of  early evolution stages of the universe.   The inflation era and  subsequent stages of evolution of the universe played an important role on the spectrum of the  waves.  The reheating stage  of the universe after inflation era  is supposed  to be an  important evolution stage of the universe.  The  relic gravitational waves being used  to determine the reheating temperature of the universe. Actually the reheating stage is model dependent but later a stage called $z$ is included to consider a general reheating scenario in the context of relic gravitational waves.   On the consideration of end of the inflationary stage and thermo nuclear synthesis, the reheating temperature   is more than a few MeV and  less than that  10$^{16}$ GeV.  

It is shown that the reheating stage and reheating temperature   can
be affect on the shape of the spectrum of the gravitational waves based on the WMAP-9 and Planck data. Therefore there is no reason for looking for detection of gravitational waves in the high range. The compared results between  the sensitivity of LISA, BBO, ultimate-DECIGO and spectrum of the waves show that all detectors can detect the waves
corresponds to the amount of $T_{rh}$. But, there
is no chance for detection of the waves due to minimum
amount of $T_{rh}$ with  LISA and BBO while it exists for the maximum amount of $T_{rh}$
especially with the ultimate-DECIGO in Fig.[\ref{ff1}]. Also  there is some more chance for detection of the waves
in full range of the $T_{rh}$ with ultimate-DECIGO in Fig.[\ref{ff2}]. Hence the future mission of the mentioned detectors especially ultimate-DECIGO are likely to detect  the waves.

\appendix \label{m}
 \section {}
The tensor to scalar ratio is $r=P_{T}(k)/P_{S}(k)$,
where the tensor power spectrum $P_{T}(k)$ and scalar power spectrum $ P_{S}(k)$  are as follows
  
\begin{equation}
P_{T}(k)=P_{T}(k_{0}^{p})(\dfrac{k}{k_{0}^{p}})^{n_{t}(k_{0}^{p})+\dfrac{1}{2}\alpha_{t}\ln(k/k_{0}^{p})},
\end{equation}

\begin{equation}
P_{S}(k)=P_{S}(k_{0}^{p})(\dfrac{k}{k_{0}^{p}})^{n_{s}(k_{0}^{p})-1+\dfrac{1}{2}\alpha_{s}\ln(k/k_{0}^{p})}.
\end{equation}
The parameters $n_{t}, n_{s}$ are tensor and scalar spectral indexes respectively with their corresponding running $\alpha_{t}\equiv dn_{t}/d \ln k$ and $\alpha_{s}\equiv dn_{s}/d \ln k$ and also $k_{0}^{p}=0.002$ Mpc$^{-1}$ is pivot wave number \cite{adb}.

\end{document}